\documentclass[aps,showpacs,twocolumn]{revtex4}
\usepackage{epsfig,amssymb,bm,graphics,color}
\newcommand*{\fs}[1]{#1\!\!\!/}
\newcommand*{\fsk}{k\!\!\!/}
\newcommand*{\ee}{e^+e^-}
\begin{document}{\normalsize }

\title{Enhanced subthreshold ${\bf e^+ e^-}$ production in short laser pulses}

\author{A.I.~Titov$^{a,b,c}$,  H.~Takabe$^{a}$,  B.~K\"ampfer$^{c,d}$, and
A.~Hosaka$^e$}
 \affiliation{
 $^a$Institute of Laser Engineering, Yamada-oka, Suita, Osaka 565-0871, Japan\\
 $^b$Bogoliubov Laboratory of Theoretical Physics, JINR, Dubna 141980, Russia\\
 $^c$Helmholtz-Zentrum  Dresden-Rossendorf, 01314 Dresden, Germany\\
 $^d$Institut f\"ur Theoretische Physik, TU~Dresden, 01062 Dresden, Germany\\
 $^e$Research Center of Nuclear Physics, 10-1 Mihogaoka Ibaraki,
 567-0047 Osaka, Japan}

\begin{abstract}

  The emission of $\ee$ pairs off a probe photon propagating through a polarized
  short-pulsed electromagnetic (e.g.\ laser) wave field is analyzed.
  A significant increase of the total cross section of
  pair production in the subthreshold region is found for decreasing laser pulse duration
  even in case of moderate laser pulse intensities.

\end{abstract}


\maketitle

The history of the study of $\ee$ production in $\gamma' \gamma$
interaction starts with the pioneering work by Breit and
Wheeler~\cite{Breit-Wheeler-1934} published in 1934. About thirty
years later, Reiss~\cite{Reiss} and Narozhnyi, Nikishov and Ritus
\cite{NR-64,Ritus-79} have analyzed the $\ee$ emission off a
photon $\gamma'$ propagating in the field of an intensive
polarized monochromatic electromagnetic (e.m.) plane. The $\ee$
production probabilities were found using the non-perturbative
Volkov solutions for the electron and positron wave
functions~\cite{Volkov}.

 If one identifies the external e.m.\ field with a laser
 pulse then most of the early work considers long lasting pulses
 where the temporal shape can be neglected.
 We denote this approach as the infinite pulse
 approximation (IPA).
 In IPA, electrons $e^-$ and positrons $e^+$
 become quasi-particles with
 effective quasi-momenta and effective (dressed) masses.
 Differential and total probabilities of the $\ee$ pair emission
 depend on the reduced strength of the e.m.\ field $A^\mu$,
 $\xi^2=-\frac{e^2 \langle A^2  \rangle }{M^2_e}\equiv
 \frac{e^2a^2}{M_e^2}$,
 where $M_e$ is the electron mass (we use $c=\hbar=1$, $e^2/4\pi
 =\alpha=1/137$). Furthermore, the dimensionless variable
 $\zeta=\frac{s_{\rm thr}}{s}$
 is introduced,
 where $s$ is the  square of the total energy in the center of mass
 system (c.m.s.) of the Breit-Wheeler process $\gamma'+ \gamma \to e^+ + e^-$ and
 $s_{\rm thr}=4M_e^2$ is its threshold value. The Ritus variable
 is then defined by
 $\kappa=2\xi/\zeta$ \cite{Ritus-79}.  The case of $\zeta>1$
 corresponds essentially to multi-photon processes. Within IPA,
 the minimum number of photons $\gamma$ in the reaction $\gamma'+n\gamma\to e^+ + e^-$
 is defined as $n_{\min}=I(\zeta) +1$, where $I(\zeta)$ is the integer
 part of $\zeta$. First evidence of the multi-photon Breit-Wheeler
 process  with $\zeta=3.83$ and $0.1<\xi<0.35$
 was detected at SLAC in the E-144 experiment \cite{SLAC-1997}, where
the application of IPA is justified
 since the used laser pulses contain around $10^3$ cycles in a shot.

 The rapidly evolving laser technology \cite{Tadjima} can
 provide the laser power up to $10^{24}$ -- $10^{25}$~W/cm$^2$ in
 near future which is sufficient for the formation of positrons
 from cascade processes in the photon-electron-positron
 plasma~\cite{Fedotov-2010,Nerush-2011,Elkina-2011} generated by
 photon-laser~\cite{Ilderton-2011,Heinzl-2010,Denisenko-2008},
 electron-laser~\cite{Ilderton-2010,Sokolov-2010} or laser-laser
 interactions~\cite{Bell-2008,Kirk-2009} (see \cite{Bulanov-2011}
 for surveys).
 The next generation of optical laser beams are expected to be essentially
 short (femtosecond duration) with only
 a few oscillation of the e.m.\ field in the pulse to be expected
 at ELI~\cite{ELI} and CLF~\cite{CLF} facilities.
 {This requires the generalization of the
 IPA multi-photon process $\gamma'+n\gamma\to e^+ + e^-$
 to a finite pulse duration.}
 Formally, this generalization may be
 done in a straightforward manner by substituting
 the expansion in Fourier series into Fourier integrals with taking
 into account the Volkov solution for the finite wave field.
 In practice, an evaluation of the total cross section
 requires the calculation of five-dimensional integrals
 with rapidly oscillating integrands which is rather demanding.
 Therefore, previous considerations are often restricted
 to the analysis of the three-dimensional differential cross
 sections,
 see for example~\cite{Heinzl-2010} for
 finite beam size effects in $\ee$ pair production
 (cf.\ also \cite{Hebenstreit-2011} and references therein).

 The aim of the present Letter is to elaborate a method for the
 calculation of the total cross section in the subthreshold
 (multi-photon) region
 accounting for the effect of finite laser pulse duration in $\ee$ pair
 production off a probe photon.
 We denote such a process with a
 finite pulse and plane wave fronts as finite pulse approximation (FPA).
 In this case,
 the in/out fermion states refer to the vacuum. Moreover, due to
 the modulation of the pulse envelope function, the power spectrum
 contains frequencies $> \omega$ (see below) which enhance the pair
 production in the subthreshold region even for moderately strong
 laser intensities.

 We consider the e.m.\ four-potential $A \sim (0,{\bf A})$
 in FPA, depending solely on the invariant phase $\phi = k \cdot x$,
 \begin{eqnarray} {\bf A}(\phi) =   f(\phi) 
( {\bf a}_1\cos\phi+{\bf a}_2\sin\phi) ~,
 \label{III1}
 \end{eqnarray}
 where $|{\bf a}_1|=|{\bf a}_2|=a,\,\,{\bf a}_1{\bf a}_2=0$ for
 circular polarization. We employ here the envelope function
 $f(\phi)= 1 / \cosh(\phi/\Delta)$, where
 $\Delta=\pi\frac{\tau}{\tau_0}=\pi N$, and $N$ characterizes the
 number of cycles in a pulse; $\tau_0 = 2\pi / \omega$ is the
 time of one cycle for the laser frequency $\omega$. Thus, $\tau$
 is the time scale of the pulse duration. The case of pulses
 obeying $\omega \tau \gg 1$ has been analyzed in
 \cite{Narozhny_Fofanov}.

 Utilizing the e.m.\ potential (\ref{III1}) in the Volkov solutions leads to two
 significant modifications of the transition amplitude. Besides
 physical asymptotic momenta and masses, the finite time $\tau$ requires
 Fourier integrals in the integrand of invariant amplitudes,
 and the discrete harmonics become continuous. Thus, the $S$ matrix
 element of the process
 $\gamma' \to e^+ (\gamma) + e^- (\gamma)$,
 where $e^\pm (\gamma)$ refers to Volkov states in the field  (\ref{III1}),
 is expressed as
 \begin{eqnarray}
 S = \int_\zeta^\infty
 dl \, M(l)\frac{(2\pi)^4\delta^4(k'+ lk -p-p')}{\sqrt{2p_02p_0'2\omega'}} ,
 \label{III3}
 \end{eqnarray}
 where the transition matrix $ M(l)$, similarly to
 the case of the non-linear Compton effect
 \cite{Boca-2009,Mackenroth-2011,NF-96,Seipt-2011} as a crossed channel
 of the pair production, consists of four terms
 \begin{eqnarray}
 M(l)=\sum_{m=0}^3  M^{(m)}\,C^{(m)}(l)~,
 \label{III4}
 \end{eqnarray}
 where
 \begin{eqnarray}
 C^{(m)}(l)&=&\frac{1}{2\pi}\int_{-\infty}^{\infty} d\phi \
 \chi^{(m)}(\phi)\, \,{\rm e}^{i l \phi -i{\cal P(\phi)}}~.
 \label{III5}
 \end{eqnarray}
 Here, $\chi^{(m)} = (1, f^2 (\phi), f(\phi) \cos \phi, f(\phi)
 \sin \phi)$ with $m = 0, 1, 2, 3$ and
 \begin{eqnarray}
 {\cal P(\phi)}&=&z{\cal P}_0(\phi,\phi_0)-\xi^2\zeta u
 \int_{-\infty}^\phi d\phi' \,f^2(\phi') ,
 \label{III6} \\
 {\cal P}_0(\phi,\phi_0) &=& \int_{-\infty}^{\phi} d\phi'\,
 \cos(\phi'-\phi_0)f(\phi') ,
 \label{III6_}
 \end{eqnarray}
 where $u=(k \cdot k')^2/(4(k \cdot p)(k \cdot p'))$, $z=2l\xi\sqrt{u(u_0-u)}/u_0$,
 $u_0=l/\zeta$. The angle $\phi_0$ is related to the azimuthal
 angle of the positron in the $\ee$ rest frame by
 $\phi_0=\phi_p+\pi$ and can be determined through invariants
 $\alpha_{1,2}=e\left((a_{1,2}\cdot p)/(k\cdot p)-(a_{1,2}\cdot p')/(k\cdot p')\right)$
 as $\cos\phi_0=\alpha_1/z$, $\sin\phi_0=\alpha_2/z$. Here,
 $p_{e^-} \equiv p' \sim (p_0', {\bf p}')$ and
 $p_{e^+}\equiv p \sim(p_0, {\bf p})$. The transition operators
 $M^{(2,3)}$ are the same as in IPA~\cite{Ritus-79}, while the
 operators $M^{(0,1)}=\bar u_{p'}\, \hat M^{(0,1)} \,v_p$, read now
 \begin{eqnarray}
 \hat M^{(0)}=\fs\varepsilon'\, ,
 \,\,\hat M^{(1)}=\frac{e^2{\fs A}\,\fsk\fs\varepsilon'\,\fsk\fs{ A}}
 {4(k\cdot p)(k\cdot p')},
 \label{III7}
 \end{eqnarray}
 where $u_{p'}$ and $v_{p}$ are the free-field Dirac spinors of the
 outgoing electron and positron, respectively; $\varepsilon'$ is
 the polarization four-vector of the probe photon $\gamma'$ with four-momentum
 $k' \sim (\omega', {\bf k}')$, and $k \sim (\omega, {\bf k})$
 is the four-momentum of the e.m.\ (laser) field
 (\ref{III1}).
 Feynman's slash notation is employed, e.g.\ ${\fs A} = A \cdot \gamma$,
 as four-product with the
 Dirac $\gamma$ matrices. 
 The integrand of the function $C^{(0)}$ does not contain the envelope
 function and needs a regularization, e.g.\ using a prescription
 given in  Ref.~\cite{Boca-2009}
 \begin{eqnarray}
 C^{(0)}(l)&=&\frac{1}{2\pi l}\int_{-\infty}^{\infty} d\phi
 \,{\rm e}^{il\phi -i{\cal P(\phi)}} \nonumber\\
 &\times& \left( z\cos(\phi-\phi_0)\,f(\phi)
 -\xi^2 {\zeta} u\,f^2(\phi)\right)~.
 \label{III8}
 \end{eqnarray}

 The probability is normalized to some time unit.
 In IPA, one can use the time of one cycle, $\tau_0$.
 In FPA, a proper time unit is provided by the pulse width,
 which is $N$ times greater, $\tau=N\tau_0$, where $N$ is the number of
 the cycles in a pulse. Therefore, for a convenient comparison of IFA and FPA results,
 the latter one is scaled by $1/N$. Thus,
 the probability of the $\ee$ pair emission reads
 \begin{eqnarray}
 W^{FPA}=\frac{\alpha M_e^2}{4\omega'N}
 \int\frac{d\phi_p}{2\pi}
 \int_\zeta^\infty dl
 \int_1^{u_0}du
  \, \frac{w(l,\xi,u,\phi_p)}{u^{3/2}\sqrt{u-1}},
  \label{III9}
 \end{eqnarray}
 \begin{eqnarray}
 && w(l,\xi,u,\phi_p)= (2u_0+1)|C^{(0)}(l)|^2 \label{III20} \\
 &&+\xi^2(2u-1)
 \left(|C^{(2)}(l)|^2 +|C^{(3)}(l)|^2 \right) \nonumber \\
 &&+  {\rm Re}\,
 C^{(0)}(l)\left( \xi^2 {C^{(1)}(l)}
 - \frac{2}{\zeta}[\alpha_1 {C^{(2)}(l)} +\alpha_2 C^{(3)}(l)] \right)^* \nonumber
 \end{eqnarray}
 with $u_0=l/\zeta$. This expression will be used below for direct numerical evaluations of
 the probability.

 Inspection of the functions ${\cal P}(\phi)$ and
 $C^{(m)}(l)$ shows however that Eq.~(\ref{III20}) may be
 simplified to get, in some cases, a more suitable analytical
 expression for $w(l)$. Integrating by parts, the function ${\cal
 P}_0(\phi,\phi_0)$ might be expressed in the following form
 \begin{equation}
{\cal P}_0 (\phi,\phi_0) = \sin(\phi-\phi_0)f(\phi) + {\cal O}(\Delta) ,
\end{equation}
where ${\cal O}(\Delta) =
-\frac{1}{\Delta}\int_{-\infty}^\phi\,\sin(\phi'-\phi_0)f'(\phi')d\phi'$
is a rather small contribution for
a finite pulse duration $\Delta=\pi N$ with $N\ge2$ because
 of (i) the factor $1/\Delta$ and (ii) the
 derivative $f'(\phi)$ in the integrand has a maximum value at the
 boundaries of the pulse with $\phi\sim 0.9\Delta$, where this
 function is suppressed.
In fact, the numerical evaluation shows that the contribution of
${\cal O}(\Delta)$ can be omitted (we find $\vert {\cal O}(\Delta)\vert < 0.1$
(0.05) for $\Delta = 2 \pi$ ($5 \pi$)). This
approximation allows to express the basic functions $C^{(m)}(l)$
via new functions $Y_l$ and $X_l$
\begin{eqnarray}
 C^{(0)}(l)&=&\widetilde Y_l(z)\,{\rm e}^{il\phi_0},\,\,\,\,
 C^{(1)}(l)=X_l(z)\,{\rm e}^{il\phi_0}~,  \label{III25}\\
 C^{(2)}(l)&=&\frac{1}{2}\left( Y_{l+1}(z)\,{\rm e}^{i(l+1)\phi_0}
 + Y_{l-1}(z)\,{\rm e}^{i(l-1)\phi_0}\right)~,\nonumber
\end{eqnarray}
with
\begin{eqnarray}
 \widetilde Y_l(z)&=&\frac{z}{2l} \left(Y_{l+1}(z) +
 Y_{l-1}(z)\right) -\xi^2u\frac{\zeta}{l}\,X_l(z)~, \nonumber \\
Y_l(z)&=&\frac{1}{2\pi} \int_{-\infty}^{\infty}\,
d\psi\,\tilde{f}^{(1)}(\psi + \phi_0)
\,{\rm e}^{il\psi-izf(\psi+\phi_0)\sin\psi}~,\nonumber\\
X_l(z)&=&\frac{1}{2\pi} \int_{-\infty}^{\infty}\,
d\psi\,\tilde{f^{(2)}}(\psi + \phi_0)
\,{\rm e}^{il\psi-iz f(\psi+\phi_0)\sin\psi}~,\nonumber\\
\tilde{f}^{(m)}(\phi)&=&f^m(\phi)\,\exp[i\xi^2\zeta\,u\,\tanh\frac{\phi}{\Delta}]~.
\label{III24}
\end{eqnarray}
 The function $C^{(3)}$ emerges from $C^{(2)}$ by
 the substitutions $1/2\to1/2i$ and sign $"+"$
 between two terms in the bracket to $"-"$.
 In the last line, "$m$" (=1, 2) is a label on the l.h.s.,
 while on the r.h.s.\ it is
 the power of the envelope function,
 as follows from Eqs.~(\ref{III5}), (\ref{III8});
 the exponential term results from an analytic evaluation of
 the last term in Eq.~(5) for the chosen envelope function.

The partial probability $w(l)$ in Eq.~(\ref{III20}) reads
\begin{eqnarray}
&& w(l,\xi, u,\phi_p)= 2 \widetilde Y^2_l(z) +\xi^2(2u-1) \nonumber\\
 &&\times\left(Y^2_{l-1}(z)+
 Y^2_{l+1}(z)-2{\rm Re}\,\widetilde Y_l(z)X^*_l(z)\right)
 \label{III26}
\end{eqnarray}
 which resembles the expression for the probabilities $w_n$ in case of IPA
 (cf.\ Ref.~\cite{Ritus-79}) arising upon the substitutions
 $\int\,dl w(l) \to \sum_n w_n$,
 $\widetilde Y^2_l \to J_n^2$,
 $Y^2_{l\pm1} \to J^2_{n\pm1}$,
 ${\rm Re}\,\widetilde Y_l(z)X^*_l(z)
 \to J^2_n$ with Bessel functions $J_n$.

In the case of small field intensity, $\xi\ll1$, implying $z\ll
1$, and denoting $l=n +\epsilon$, where $n$ is the integer part of
$l$, one can use the following decomposition
\begin{eqnarray}
&&Y_l\simeq\frac{1}{2\pi}\int_{-\infty}^{\infty}\,d\psi
\,{\rm e}^{il\psi -izf(\psi+\phi_0)\sin\psi} f(\psi+\phi_0)\label{B5}\\
&&\to \frac{1}{2\pi}\int_{-\infty}^{\infty}\,d\psi
\sum_{k=0}^{\infty} \frac{(iz)^k}{k!}\sin^k\psi \,{\rm
e}^{i(n+\epsilon)\psi} f^{k+1}(\psi+\phi_0) \nonumber
\end{eqnarray}
and analog for the function $X_l(z)$ with the substitution
$f^{k+1}\to f^{k+2}$. The dominant contribution to the integral
with a rapidly oscillating integrand stems from the term with
$k=n$, which results in
\begin{eqnarray}
Y_{k+\epsilon}\simeq \frac{z^k}{2^kk!} \, {\rm
e}^{-i\epsilon\phi_0}
f_F^{(k+1)}(\epsilon),\nonumber\\
X_{k+\epsilon}\simeq \frac{z^k}{2^kk!} \, {\rm
e}^{-i\epsilon\phi_0} f_F^{(k+2)}(\epsilon), \label{B6}
\end{eqnarray}
where the function $f_F^{(k)}(\epsilon)$ is the Fourier transform of
the function $f^k(\psi)$.
For the above envelope function 
it can be calculated analytically using the theory of residues.
Results of the leading orders $n=0,1$ are
\begin{eqnarray}
Y_{0+\epsilon}(z)&=& \frac{\Delta \,{\rm
e}^{-\pi|\epsilon|\Delta/2}} {1+ {\rm e}^{-\pi|\epsilon|\Delta}}
\,{\rm e}^{-i\epsilon\phi_0}~,\nonumber\\
Y_{1+\epsilon}(z)&=&\frac{z}{2} \frac{\Delta^2|\epsilon|\, {\rm
e}^{-\pi|\epsilon|\Delta/2}} {1- {\rm e}^{-\pi|\epsilon|\Delta}}
\,{\rm e}^{-i\alpha\phi_0}~,\nonumber\\
X_{1+\epsilon}(z)&=&\frac{z}{4}
\frac{\Delta(\Delta^2\epsilon^2+1)\,{\rm
e}^{-\pi|\epsilon|\Delta/2}} {1 + {\rm e}^{-\pi|\epsilon|\Delta}}
\,{\rm e}^{-i\epsilon\phi_0}~. \label{B8}
\end{eqnarray}
 The representation of Eq.~(\ref{B6}) evidences (i) a fast decrease
 of $Y^2_{n+\epsilon}$ with increasing $|\epsilon|$
 and (ii) the $\phi_p$ dependence disappears
 in $Y^2_{n+\epsilon}$ and $X^2_{n+\epsilon}$. This allows to
 express the integral over $dl$ in (\ref{III9}) in a form useful
 for a qualitative analysis:
\begin{eqnarray}
&\left(\frac{4\omega' N}{\alpha M_e^2}\right)& W^{FPA}
 =
 \int_{\zeta-n_{0}}^1 d\epsilon
 \int_1^{u_0} du\,
 \frac{w(n=n_{0},\epsilon,\xi,u)}{u^{3/2}\sqrt{u-1}} \nonumber\\
 &&+\sum\limits_{n=n_{0}+1}^{\infty}
 \int_{\nu}^1 d\epsilon
  \int_1^{u_0}du\,
 \frac{w(n, \epsilon,\xi,u)}{u^{3/2}\sqrt{u-1}}
 \label{integral}
\end{eqnarray}
 with $u_0=(n+\epsilon)/\zeta$; $n_{0}=1$ for $\zeta \le1$, and
 $n_0=I(\zeta)$ for $\zeta >1$;
 The lower limit in integral over $d\epsilon$ in the second term reads
 $\nu=\zeta-n$ for
 $\zeta>1$ and $n=n_0+1$, and $\nu=-1$ in other cases.
 This equation shows that, contrary to IPA where at given $\zeta>1$
 (i.e.\ below threshold, $s < s_{\rm thr}=4M_e^2$) only harmonics
 with $n>I(\zeta+1)$ contribute, in FPA the harmonic with
 $n=I(\zeta)$ also contributes.

 Consider, as a check of the normalization,
 the pair production above threshold with $\zeta=1-\delta s_e/s < 1$,
 where $\delta s_e$ is the energy excess $\delta s_e=s-s_{\rm thr}$.
 Utilizing the explicit expressions~(\ref{B8})
 for the leading contribution $\widetilde Y_{1+\epsilon}$ in (\ref{III26})
 one can get a relation
 between emission probabilities in IPA (cf.~\cite{Ritus-79}) and
 FPA:
 \begin{eqnarray}
 && W^{\rm FPA} = W^{\rm IPA}(n=1,\xi,\bar u_1)\, {\cal I}(\Delta,\zeta),
 \label{III28}\\
 && {\cal I}(\Delta,\zeta) = \frac{\Delta^2}{N} \int_{\zeta-1}^{1}
 d \epsilon \frac{{\rm e}^{-\pi|\epsilon|\Delta}}
 {(1+ {\rm e}^{-\pi|\epsilon|\Delta})^2}~,  
 \end{eqnarray}
 where $\bar u_n$ is an effective value of $u$ in $n$th
 term of Eq.~(\ref{integral}). The dependence of $W$ on $\bar u_n$ is rather
 weak compared to the dependence on $\xi$ and can be disregarded.
 Thus, in the limit $\pi\Delta\delta s_e/s\gg1$,
 IPA and FPA practically coincide since
 ${\cal I}(\Delta,\zeta)  \simeq {\Delta}
 [\pi N(1+  {\rm e}^{-\pi\Delta\delta s_e/s})]^{-1}
 \simeq \frac{\Delta}{\pi N}=1$.

 Consider now the case of subthreshold pair production with
 $\zeta=1+\delta s_l/s > 1$, where $\delta s_l=s_{\rm thr}-s$
 is the "lack of energy". The probability has the following form
 \begin{equation}  \label{III30-1}
 W^{\rm FPA} = {\cal I}_1 W^{\rm IPA}(n=1) +
 C W^{\rm IPA}(n=2)+\,...,
\end{equation}
 with ${\cal I}_1(\Delta,\delta s_l/s) \simeq { {{\rm e}^{-\pi\Delta
 \delta s_l/s}}} /{({1+ {\rm e}^{-\pi\Delta\delta s_l/s}})}$
 and $C=(1/\pi^2)\int_{\pi\Delta(\zeta-2)}^{\pi\Delta}
 x^2\exp(-x) (1-\exp(-x))^{-2} dx \simeq 2/3$ for
 $\delta s_l/s \lesssim 1 - 0.65/N$.
 The terms in r.h.s.\ of (\ref{III30-1}) are meant to have the
 same functional dependence on $\xi$ and  $\bar u_{1,2}$ as in IPA.
 One can expect a significant enhancement of pair production for the
 short pulse because the probability of single
 photon events ($n=1$) is much greater than the probability
 of the two-photon events ($n=2)$:
 $W^{\rm IPA}(n=1)/W^{\rm IPA}(n=2) \sim \xi^{-2} \gg 1$. When
 the length of the pulse increases the contribution of the first term in
 Eq.~(\ref{III30-1}) decreases exponentially due to ${\cal I}_1$,
 and the prediction of FPA approaches to the IPA one.

\begin{figure}[h!]
\includegraphics[width=0.45\columnwidth]{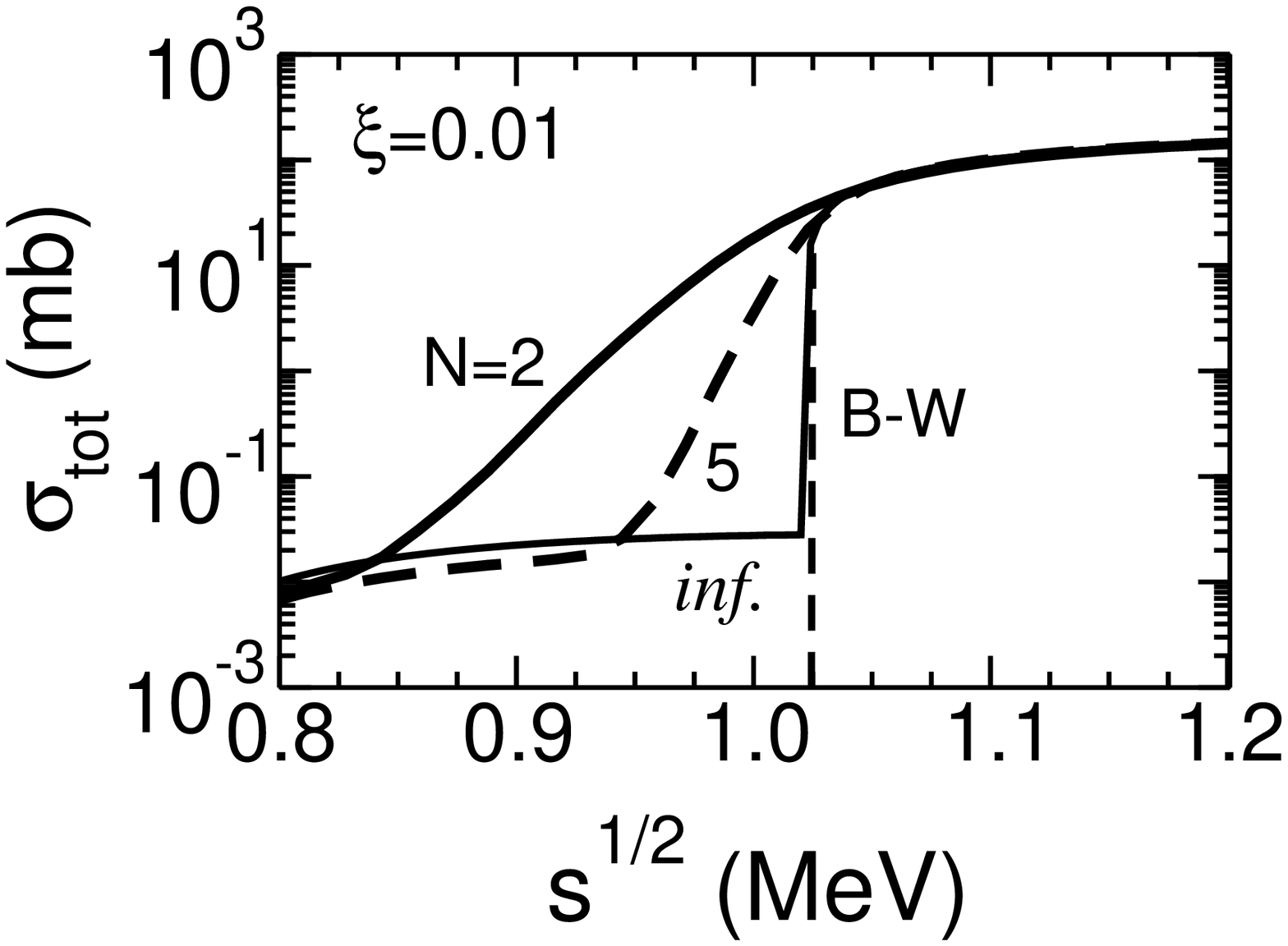}
\qquad
\includegraphics[width=0.45\columnwidth]{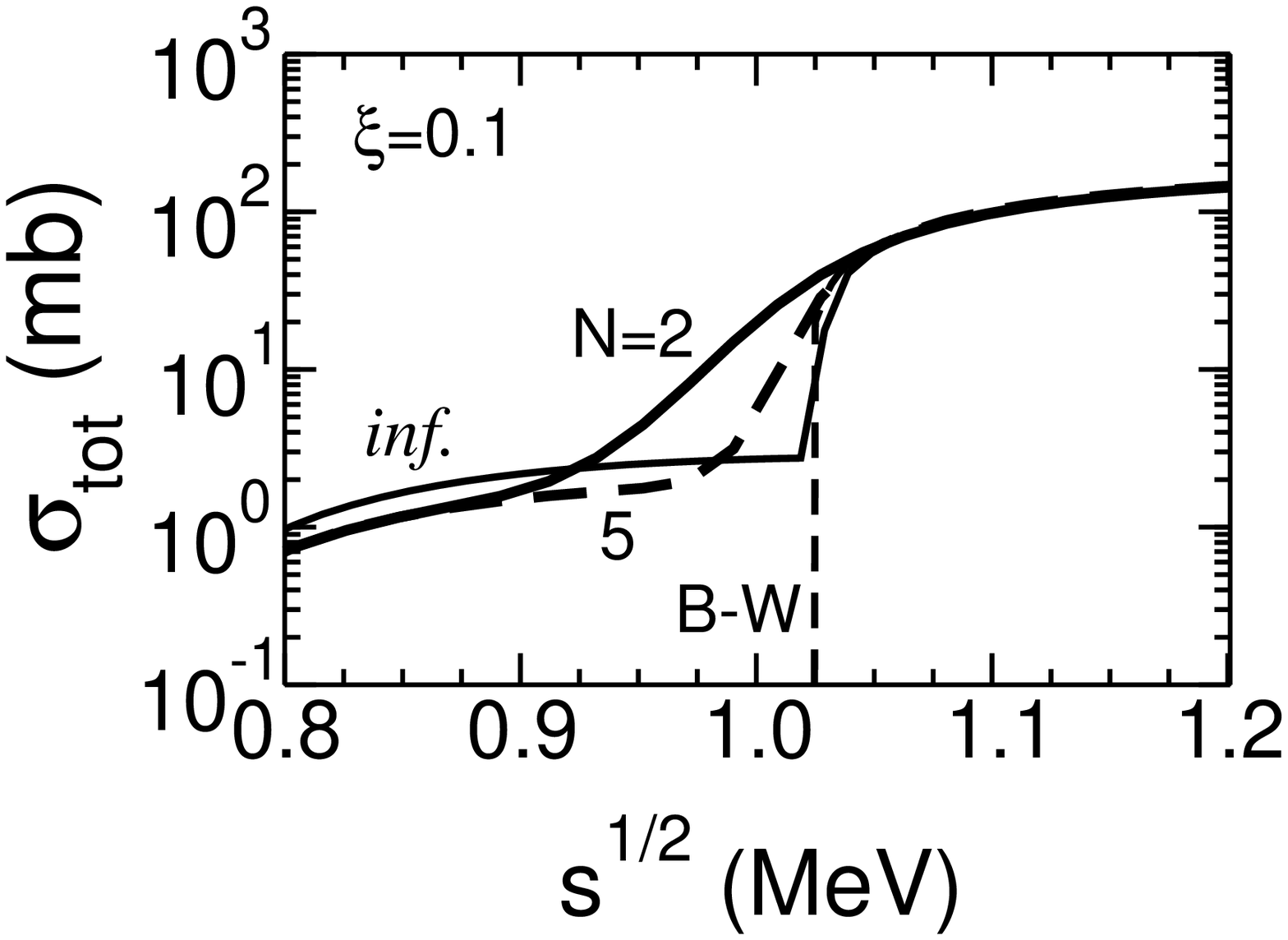}
\caption{\small{ The total cross section of $\ee$ pair production
as a function of the total energy in c.m.s., $\sqrt{s}$, for the
finite pulse. Notation is given in the text.
%
 \label{Fig:3} }}
\end{figure}

 The probability and the cross section are related
 to each other \cite{Titov-2011} as
 $dW=2[\omega M_e^2\xi^2/(4\pi\alpha)]d\sigma $.
 The total cross section of $\ee$ production is
 calculated using Eqs.~(\ref{III9}), (\ref{III26}) and (\ref{B6}).
 The cross sections are exhibited in Fig.~\ref{Fig:3}
 as a function of $\sqrt{s}$ in the
 threshold region for finite pulses with $\Delta=\pi N$.
 The left and right panels correspond to $\xi=0.01$ and 0.1,
 respectively. The dashed and thick solid curves are for
 $N=2$ and 5, respectively. The thin solid curve is the IPA result.
 The thin dashed curve, labelled by "B-W", corresponds to the
 Breit-Wheeler process \cite{Breit-Wheeler-1934} practically
 coinciding with the lowest harmonic ($n=1$).
 One can see that in
 the subthreshold region, $\sqrt{s}=0.85$ -- $1.02$~MeV, the cross
 section for short pulses is significantly greater than in IPA
 and the difference may reach one or two orders of magnitude for $\xi=0.1$ and
 $\xi=0.01$, respectively.
 When $\xi$ and/or $\zeta$ increase, the
 contribution of higher terms with $n\ge1$ becomes finite that brings
 an additional (increasing) dependence on $\Delta$ (cf.\
 Eq.~(\ref{B6}).
\begin{figure}[h!]
\includegraphics[width=0.45\columnwidth]{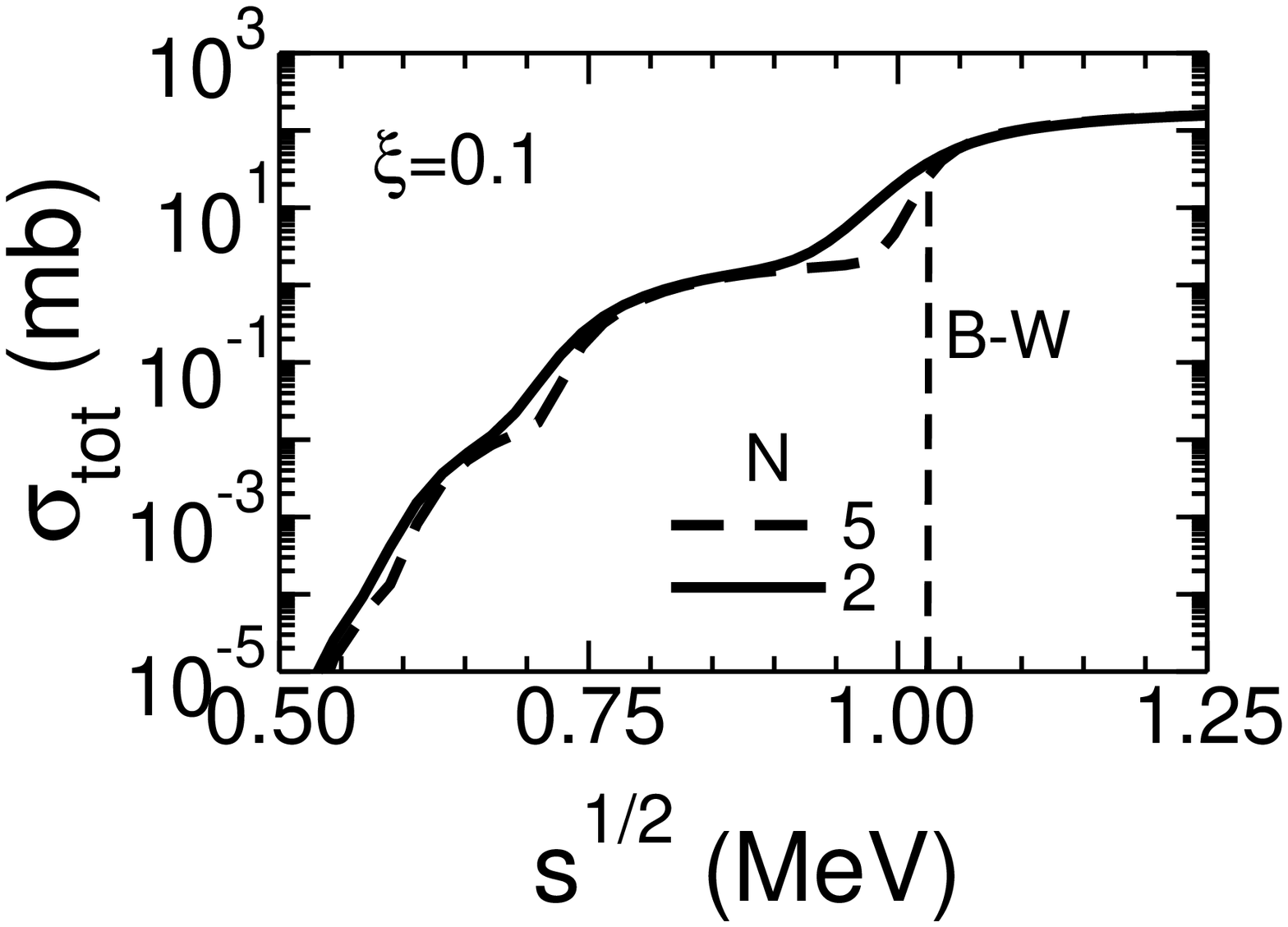}\qquad
\includegraphics[width=0.45\columnwidth]{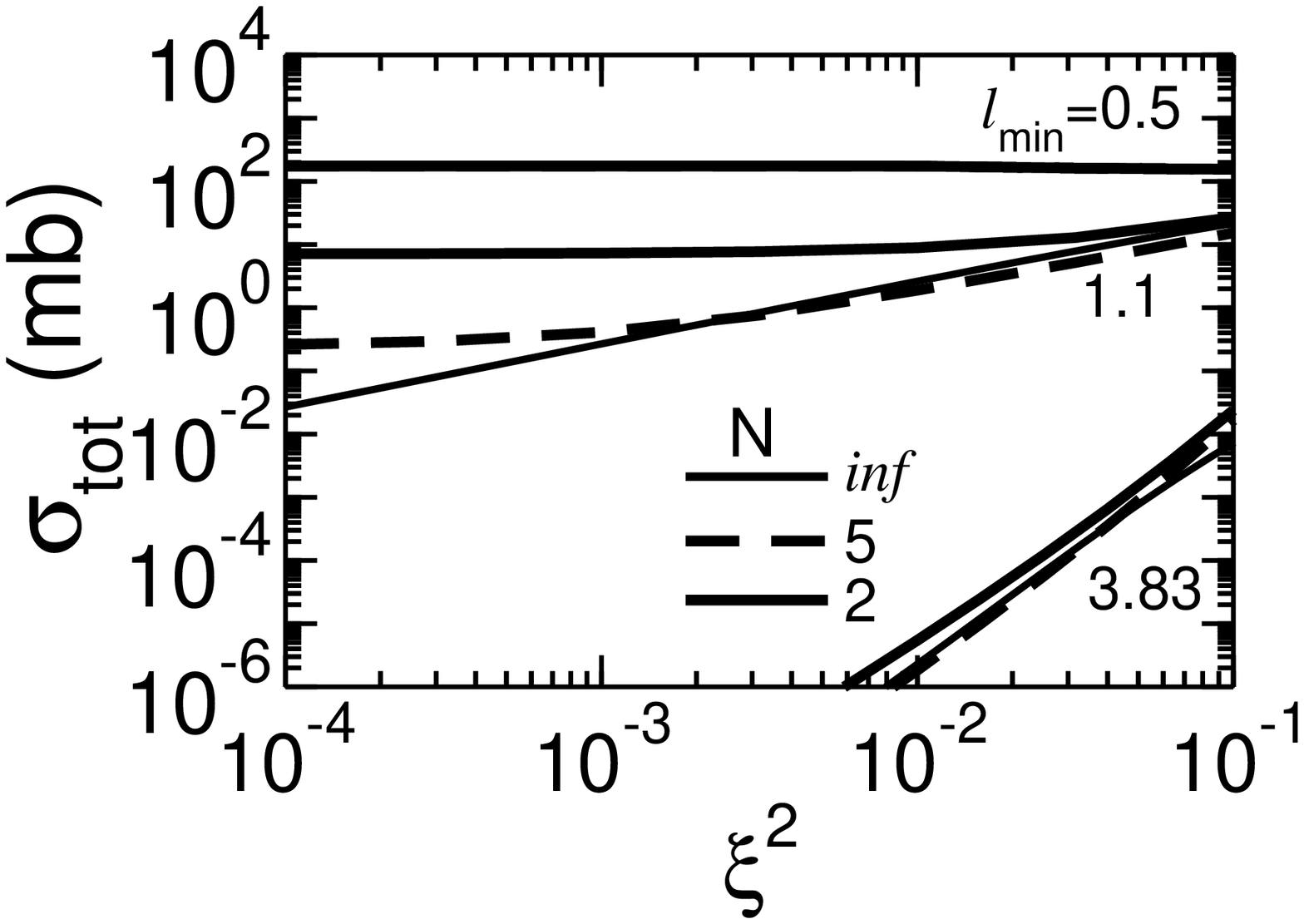}
 \caption{\small{Left panel:
As Fig.~\ref{Fig:3} (right panel) but
 for a wider region of $\sqrt{s}$.
 Right panel: The total cross section of the $\ee$-pair
 production as a function of $\xi^2$ at three values of
 $l_{\rm min} = \zeta=0.5$, 1.1 and 3.8. Notations as in Fig.~\ref{Fig:3}.
 \label{Fig:4} }}
\end{figure}

 The total cross section in a wider region of $\sqrt{s}$ is exhibited
 in Fig.~\ref{Fig:4}, left panel. At $\sqrt{s}\simeq0.55$~MeV the multi-photon
 events with $l\ge4$ become important. In general, the total cross
 section in FPA has also the step-like structures similar to IPA.
 However, a decrease of the pulse duration leads to a smoothing.
 One can also see some enhancement
 of the cross section for a short pulse with $N=2$ compared to the case
 of a longer pulse with $N=5$.
 The total cross sections of the $\ee$ pair
 production as a function of $\xi^2$ at three values of
 $l_{\rm min}=\zeta=s_{\rm thr}/s$ are presented in Fig.~\ref{Fig:4}, right panel.
 The case of $\zeta=0.5$ corresponds to the production above the threshold.
 Here, the predictions for IPA and FPA coincide.
 Examples of $\zeta=1.1$ and 3.8 correspond to the subthreshold production.
 In the first case, we are slightly below the threshold
 and one can see a large difference
 between predictions for pulses with $N=2$ and 5, which has been
 explained above.
 The last example ($\zeta=3.83$) corresponds to the kinematics of the SLAC E-144
 experiment. In this case the predictions of IPA and FPA are
 qualitatively similar with some enhancement for a shorter pulse duration.
 Finally note that we do not take into account radiation reaction
 effect discussed in~\cite{Ritus-79} and recently in Ref.~\cite{RR}
 because it influences the fermions in the final state and is
 not expected to change significantly the total $\ee$ yield.

 In summary, we have considered the total cross section of $\ee$
 production off a probe photon interacting with a semi-intensive
 short laser pulse in the subthreshold region defined by
 multi-photon interactions. We find a non-trivial dependence of the
 cross section (production probability) on the pulse duration. Just
 below the threshold of the weak-field Breit-Wheeler process, the
 short laser pulses increase the cross section up to two orders of
 magnitude relative to a monochromatic plane wave. This effect must
 be taken into account in the evaluation of $\ee$ pair production
 in cascade processes produced by high-power laser fields.

 The authors acknowledge fruitful discussions with
 T.~E. Cowan. A.I.T. appreciates ILE of Osaka University for the
 kind hospitality.

\end{document}